**Enhanced Gilbert Damping in Re doped FeCo Films – a combined experimental and theoretical study**

S. Akansel[1], A. Kumar[1], V.A. Venugopal[2], R. Banerjee[3], C. Autieri[3], R. Brucas[1], N. Behera[1], M. A. Sortica[3], D. Primetzhofer[3], S. Basu[2], M.A. Gubbins[2], B. Sanyal[3], and P. Svedlindh[1]

[1]Department of Engineering Sciences, Uppsala University, Box 534, SE-751 21 Uppsala, Sweden

[2]Seagate Technology, BT48 0BF, Londonderry, United Kingdom

[3]Department of Physics and Astronomy, Uppsala University, Box 516, SE-751 20 Uppsala, Sweden

The effects of rhenium doping in the range 0 – 10 at% on the static and dynamic magnetic properties of $Fe_{65}Co_{35}$ thin films have been studied experimentally as well as with first principles electronic structure calculations focussing on the change of the saturation magnetization ($M_s$) and the Gilbert damping parameter ($\alpha$) Both experimental and theoretical results show that $M_s$ decreases with increasing Re doping level, while at the same time $\alpha$ increases. The experimental low temperature saturation magnetic induction exhibits a 29% decrease, from 2.31T to 1.64T, in the investigated doping concentration range, which is more than predicted by the theoretical calculations. The room temperature value of the damping parameter obtained from ferromagnetic resonance measurements, correcting for extrinsic contributions to the damping, is for the undoped sample $2.7 \times 10^{-3}$, which is close to the theoretically calculated Gilbert damping parameter. With 10 at% Re doping, the damping parameter increases to $9.0 \times 10^{-3}$, which is in good agreement with the theoretical value of $7.3 \times 10^{-3}$. The increase in damping parameter with Re doping is explained by the increase in density of states at Fermi level, mostly contributed by the spin-up channel of Re. Moreover, both experimental and theoretical values for the damping parameter are observed to be weakly decreasing with decreasing temperature.

# 1. INTRODUCTION

During the last decades, thin films of soft magnetic alloys such as NiFe and FeCo have been in focus due to possible use in applications such as spin valves,[1,2] magnetic tunneling junctions,[3,4,5] spin injectors,[6] magnetic storage technologies and in particular in magnetic recording write heads.[7] Besides spintronic and magnetic memory devices, such materials are useful for shielding applications that are necessary in order to reduce the effect of electromagnetic fields created by electronic devices. The magnetic damping parameter of the material plays a critical role for the performance of such spintronic and memory devices as well as for shielding applications. On the one hand, a low damping parameter is desired in order to get low critical switching current in spintronic devices.[8,9,10] On the other hand, a high damping parameter is necessary in order to reduce the magetization switching time in magnetic memory devices and to be able to operate devices at high speeds.[11] FeCo alloys are promising materials for high frequency spintronic applications and magnetic recording devices due to their high saturation magnetization ($M_s$), high permeability, thermal stability and comparably high resistivity.[12,13,14] One possible drawback is that FeCo alloys exhibit high coercivity ($H_c$), which is not favorable for such applications, however this problem can be solved by thin film growth on suitable buffer layers.[15,16,12] Except coercivity problems, the damping parameter of these materials should be increased to make them compatible for high speed devices.

Dynamic properties of magnetic materials are highly dependent on the damping parameter. This parameter is composed of both intrinsic and extrinsic contributions. The intrinsic contribution is called the Gilbert damping and depends primarily on the spin-orbit coupling.[17] Intrinsic damping is explained as scattering of electrons by phonons and magnons.[18,19] Besides electron scattering, due to the close relation between magnetocrystalline anisotropy and spin-orbit coupling, it can be assumed that the intrinsic damping is also related to the magnetocrystalline anisotropy constant.[20] Regarding extrinsic damping, there can be a number of different contributions. The most common contribution originates from two magnon scattering (TMS).[21] However, this contribution vanishes when ferromagnetic resonance (FMR) measurements are performed by applying the static magnetic field along the film normal in inplane anisotropic thin films.[22] Besides TMS, there are some other extrinsic contributions to the damping that are not possible to get rid of by changing the measurement configuration. One of these contributions is radiative damping, which arises from inductive coupling between the precessing magnetization and the waveguide used for FMR measurements.[23] Another contribution for metallic ferromagnetic films is the eddy current damping related to microwave magnetic field induced eddy currents in the thin films during measurements.[23,24]

In order to make a soft magnetic thin film suitable for a specific application, taking into account requirements set by the device application, its damping parameter should be tailored. As mentioned above, an increased damping parameter is necesssary for devices requiring high switching speed. Several efforts have been made for enhancing the damping parameter of soft magnetic materials. NiFe alloys constitute one of the most studied systems in this respect. The most common way to enhance the intrinsic damping of an alloy is to dope it with different elements. Rare earth elements with large spin-orbit coupling have revealed promising results as dopants in terms of increased damping parameter.[25,26,27] 3d, 4d and 5d transition metals dopants have also been studied experimentally, revealing an increase of the damping parameter.[28,29] Besides experimental results, theoretical calculations support the idea that transition metals and especially 5d elements can enhance the damping parameter of NiFe alloys due to scattering in presence of chemical disorder , as well as due to the effect of spin-orbit coupling.[30]

Although NiFe alloys have been the focus in several extensive studies, FeCo alloys have so far not been studied to the same extent. Attempts have been made to dope FeCo with Yb,[20] Dy,[31] Gd,[32] and Si,[33] where in all cases an increase of the damping parameter was observed. Apart from doping of alloys, the addition of adjacent layers to NiFe and CoFe has also been studied. In particular, adding layers consisting of rare earth elements with large orbital moments gave positive results in terms of increased damping parameter.[34]

$Fe_{65}Co_{35}$ alloys are attractive materials because of high $M_s$ and reduced $H_c$ values. However, not much is known about the magnetic damping mechanisms for this composition. Since it is of interest for high data rate magnetic memory devices, the damping parameter should be increased in order to make the magnetic switching faster. To the best of our knowledge, systematic doping of $Fe_{65}Co_{35}$ with 5d elements has not been studied so far experimentally. Some of us have found from ab initio calculations that 5d transition metal dopants can increase the damping parameter and Re is one of the potential candidates.[35] Re is particularly interesting as it has a nice compromise of having not so much reduced saturation magnetization and a quite enhanced damping parameter. In this work, we have perfomed a systematic ab initio study of $Fe_{65}Co_{35}$ doped with increasing Re concentration to find an increasing damping parameter. The theoretical predictions are confirmed by results obtained from temperature dependent FMR measurements performed on Re doped $Fe_{65}Co_{35}$ films.

## 2. EXPERIMENTAL AND THEORETICAL METHODS

Rhenium doped $Fe_{65}Co_{35}$ samples were prepared by varying the Re concentration from 0 to 10.23 at%. All samples were deposited using DC magnetron sputtering on $Si/SiO_2$ substrates. First a 3 nm thick Ru seed layer was deposited on the $Si/SiO_2$ substrate followed by room temperature deposition of 20 nm and 40 nm thick Re-doped $Fe_{65}Co_{35}$ films by co-sputtering between $Fe_{65}Co_{35}$ and Re targets. Finally, a 3 nm thick Ru layer was deposited as a capping layer over the Re-doped $Fe_{65}Co_{35}$ film. The nominal Re concentration was derived from the calibrated deposition rate used in the deposition system. The nominal Re doping concentrations of the $Fe_{65}Co_{35}$ samples are as follows; 0, 2.62, 5.45 and 10.23 at%.

The crystalline structure of the fims were investigated by utilizing grazing incident X-Ray diffraction (GIXRD). The incidence angle was fixed at 1° during GIXRD measurements and a

CuKα source was used. Accurate values for film thickness and interface roughness were determined by X-ray reflectivity (XRR) measurements.

Beside XRD, composition and areal density of the films were deduced by Rutherford backscattering spectrometry[36] (RBS) with ion beams of 2 MeV $^4$He$^+$ and 10 MeV $^{12}$C$^+$. The beams were provided by a 5 MV 15SDH-2 tandem accelerator at the Tandem Laboratory at Uppsala University. The experiments were performed with the incident beam at 5° with respect to the surface normal and scattering angles of 170° and 120°. The experimental data was evaluated with the SIMNRA program.[37]

In-plane magnetic hysteresis measurments were performed using a Magnetic Property Measurement System (MPMS, Quantum Design).

Ferromagnetic resonance measurements were performed using two different techniques. First in-plane X-band (9.8GHz) cavity FMR measurements were performed. The setup is equipped with a goniometer making it possible to rotate the sample with respect to the applied magnetic field; in this way the in-plane anisotropy fields of the different samples have been determined. Besides cavity FMR studies, a setup for broadband out-of-plane FMR measurements have been utilized. For out-of-plane measurements a vector network analyzer (VNA) was used. Two ports of the VNA were connected to a coplanar waveguide (CPW) mounted on a Physical Property Measurement System (PPMS, Quantum Design) multi-function probe. The PPMS is equipped with a 9T superconducting magnet, which is needed to saturate Fe$_{65}$Co$_{35}$ films out-plane and to detect the FMR signal. The broadband FMR measurements were carried out at a fixed microwave frequency using the field-swept mode, repeating the measurement for different frequencies in the range 15 – 30GHz.

The theoretical calculations are based on spin-polarized relativistic multiple scattering theory using the Korringa-Kohn-Rostoker (KKR) formalism implemented in the spin polarized relativistic KKR code (SPR-KKR). The Perdew-Burke-Ernzerhof (PBE) exchange-correlation functional within generalized gradient approximation was used. The equilibrium lattice parameters were obtained by energy minimization for each composition. Substitutional disorder was treated within the Coherent Potential Approximation (CPA). The damping parameters were calculated by the method proposed by Mankovsky et al.,[38] based on the ab initio Green's function technique and linear response formalism where one takes into consideration scattering processes as well as spin-orbit coupling built in Dirac's relativistic formulation. The calculations of Gilbert damping parameters at finite temperatures were done using an alloy-analogy model of atomic displacements corresponding to the thermal average of the root mean square displacement at a given temperature.

3. RESULTS AND DISCUSSION

Re concentrations and layer thickness (areal densities) of the 20 nm doped films were obtained by RBS experiments. RBS employing a beam of 2 MeV He primary ions was used to deduce the areal concentration of each layer. Additional measurements with 10 MeV C probing particles permit to resolve the atomic fractions of Fe, Co and Re. The spectra for the samples with different Re concentration are shown in Fig. A1. The measured Re concentrations are 3.0±0.1 at%, 6.6±0.3 at% and 12.6±0.5 at%. Moreover, the results for Fe and Co atomic fractions show that there is no preferential replacement by Re, implying that the two elements are replaced according to their respective concentration.

Figure 1 (a) shows GIXRD spectra in the $2\theta$-range from 20º to 120º for the $Fe_{65}Co_{35}$ films with different Re concentration. Diffraction peaks corresponding to the body centered cubic $Fe_{65}Co_{35}$ structure have been indexed in the figure; no other diffraction peaks appear in the different spectra. Depending on the Re-dopant concentration shifts in the peak positions are observed, the diffraction peaks are suppressed to lower $2\theta$-values with increasing dopant concentration. The shift for the (110) peak for the different dopant concentrations is given as an inset in Fig. 1 (a). Similar shifts are observed for the other diffraction peaks. This trend in peak shift is an experimental evidence of an increasing amount of Re dopant within the deposited thin films. Since the peaks are shifted towards lower $2\theta$-values with increasing amount of Re dopant, the lattice parameter increases with increasing Re concentration.[39] Figure 1 (b) shows the experimental as well as theoretically calculated lattice parameter versus Re concentration. The qualitative agreement between theory and experiment is obtained. However, the rate of lattice parameter increase with increasing Re concentration is larger for the theoretically calculated lattice parameter. This is not unexpected as the generalized gradient approximation for the exchange-correlation potential has a tendency to overestimate the lattice parameter. Another possible explanation for the difference in lattice parameter is that the increase of the lattice parameter for the Re-doped $Fe_{65}Co_{35}$ films is held back by the compressive strain due to lattice mismatch with $Si/SiO_2/Ru$. XRR measurements revealed that the surface roughness of the $Fe_{65}Co_{35}$ films is less than 1 nm, which cannot affect static and magnetic properties drastically. Results from XRR measurements are given in table 1.

Room temperature normalized magnetization curves for the Re-doped $Fe_{65}Co_{35}$ films are shown in Fig. 2 (a). The coercivity for all films is in the range of 2mT and all films, except for the 12.6 at% Re doped film that show a slightly rounded hysteresis loop, exhibit rectangular hysteresis loops. The low value for the coercivity is expected for seed layer grown films.[15] The experimentally determined low temperature saturation magnetization together with the theoretically calculated magnetization versus Re concentration are shown in Fig. 2 (b). As expected, both experimental and theoretical results show that the saturation magnetization decreases with increasing Re concentration. A linear decrease in magnetization is observed in the theoretical calculations whereas a non-linear behavior is seen in the experimental data.

Angle resolved cavity FMR measurements were used to study the in-plane magnetic anisotropy. The angular-dependent resonant field ($H_r$) data was analyzed using the following equation,[40]

$$f = \frac{\mu_0 \gamma}{2\pi} \left[ \left\{ H_r \cos(\phi_H - \phi_M) + \frac{H_c}{2} \cos 4(\phi_M - \phi_C) + H_u \cos 2(\phi_M - \phi_u) \right\} \left\{ H_r \cos(\phi_H - \phi_M) + M_{eff} + \frac{H_c}{8}(3 + \cos 4(\phi_M - \phi_C)) + H_u \cos^2(\phi_M - \phi_u) \right\} \right]^{1/2}, \quad (1)$$

where $f$ is the cavity resonance frequency and $\gamma$ is the gyromagnetic ratio. $\phi_H$, $\phi_M$, $\phi_u$ and $\phi_C$ are the in-plane directions for the magnetic field, magnetization, uniaxial anisotropy and cubic anisotropy, respectively, with respect to the [100] direction of the Si substrate. $H_u = \frac{2K_u}{\mu_0 M_s}$ and $H_c = \frac{4K_c}{\mu_0 M_s}$ are the uniaxial and cubic anisotropy fields, where $K_u$ and $K_c$ are the uniaxial and cubic magnetic anisotropy constants, and $M_{eff}$ is the effective magnetization. Fitting parameters were limited to $M_{eff}$, $\gamma$ and $H_u$, since the $H_r$ versus $\phi_H$ curves did not give any indication of a cubic anisotropy.

Figure 3 shows $H_r$ versus $\phi_H$ extracted from the angular-dependent FMR measurements together with fits according to Eq. (1), clearly revealing dominant twofold uniaxial in-plane magnetic anisotropy. Extracted anisotropy field and effective magnetization values are given in Table 2. The results show that $H_u$ is within the accuracy of the experiment independent of the Re concentration.

Temperature dependent out-of-plane FMR measurements were performed in the temperature range 50 K to 300 K recording the complex transmission coefficient $S_{21}$. Typical field-swept results for the real and imaginary components of $S_{21}$ for the undoped and 12.6 at% Re-doped samples are shown in Fig. 4. The field-dependent $S_{21}$ data was fitted to the following set of equations,[41]

$$S_{21}(H,t) = S_{21}^0 + Dt + \frac{\chi(H)}{\tilde{\chi}_0}$$

$$\chi(H) = \frac{M_{eff}(H-M_{eff})}{(H-M_{eff})^2 - H_{eff}^2 - i\Delta H(H-M_{eff})}. \qquad (2)$$

In these equations $S_{21}^0$ corresponds to the non-magnetic contribution to the complex transmission signal, $\tilde{\chi}_0$ is an imaginary function of the microwave frequency and film thickness and $\chi(H)$ is the complex susceptibility of the magnetic film. The term $Dt$ accounts for a linear drift of the recorded $S_{21}$ signal. $M_{eff} = M_s - H_k^\perp$, where $H_k^\perp$ is the perpendicular anisotropy field and $H_{eff} = \frac{2\pi f}{\gamma \mu_0}$. The $S_{21}$ spectra were fitted to Eq. (2) in order to extract the linewidth $\Delta H$ and $H_r$ values. Fits to Eq. (2) are shown as solid lines in Fig. 4.

The experimentally measured total damping parameter ($\alpha_{total}$), including both the intrinsic contribution (Gilbert damping) and extrinsic contributions, was extracted by fitting $\Delta H$ versus frequency to the following expression, [41]

$$\mu_0 \Delta H = \frac{4\pi \alpha_{total} f}{\gamma} + \mu_0 \Delta H_0, \qquad (3)$$

where $\Delta H_0$ is the frequency independent linewidth broadening due to sample inhomogeneity. Besides $\alpha_{total}$, $M_{eff}$ can also be extracted by fitting the $H_r$ versus frequency results to the expression

$$\mu_0 H_r = \frac{2\pi f}{\gamma} + \mu_0 M_{eff} \,. \qquad (4)$$

Typical temperature dependent results for $f$ versus $H_r$ and $\Delta H$ versus $f$ are shown in Fig. 5 for the 12.6 at% Re-doped Fe$_{65}$Co$_{35}$ film. Extracted values of $M_{eff}$ at different temperatures are given in Table 3 for all samples. As expected, the results show that $M_{eff}$ decreases with increasing dopant concentration. Since $M_{eff} = M_s - H_k^{\perp}$ and the film thickness is large enough to make a possible contribution from out-of-plane anisotropy negligible one can make the justified assumption that $M_{eff} \approx M_s$. The analysis using Eqs. (2) – (4) also give values for the Landé g-factor ($\gamma = \frac{g\mu_B}{\hbar}$), yielding 2.064 and 2.075 for the undoped and 12.6 at% doped samples, respectively (similar values are obtained at all temperatures).

As indicated above, the damping parameters extracted from FMR measurements ($\alpha_{total}$) include both intrinsic and extrinsic contributions. One of the most common extrinsic contributions is TMS, which is avoided in this study by measuring FMR with the magnetic field applied out of the film plane. Except TMS, extrinsic contributions such as eddy current damping and radiative damping are expected to contribute the measured damping. In a metallic ferromagnet, which is placed on top of a CPW, precession of spin waves induces AC currents in the ferromagnetic film, thereby dissipating energy. The radiative damping has similar origin as the eddy current damping, but here the precession of the magnetization induces microwave-frequency currents in the CPW where energy is dissipated. Thus, there are two extrinsic contributions to the measured damping; one that is caused by eddy currents in the ferromagnetic film ($\alpha_{eddy}$) and another one caused by eddy currents in the CPW ($\alpha_{rad}$).[23] In order to obtain the reduced damping of the films ($\alpha_{red}$), which we expect to be close to the intrinsic damping of the films, the extrinsic contributions should be subtracted from $\alpha_{tot}$. We have neglected any contribution to the measured damping originating from spin-pumping into seed and capping layers. However, since spin-pumping in low spin-orbit coupling materials like Ru with thickness quite less than the spin-diffusion length is quite small, the assumption of negligible contribution from spin-pumping is justified. The total damping can thus be given as $\alpha_{tot} = \alpha_{red} + \alpha_{rad} + \alpha_{eddy}$.

When the precession of the magnetization is assumed to be uniform in the sample, the expression for radiative damping can be given as[23]

$$\alpha_{rad} = \frac{\eta \gamma \mu_0^2 M_s \delta l}{2 Z_0 w}, \qquad (5)$$

where $Z_0 = 50\ \Omega$ is the waveguide impedance, $w = 240\ \mu m$ is the width of the CPW center conductor, $\eta$ is a dimensionless parameter that accounts for FMR mode profile, $\delta$ is the thickness and $l$ is the length of the sample. The length of all samples were 4mm and the thickness 20nm for the undoped and 12.6 at% Re-doped films and 40nm for the 3.0 at% and 6.6 at% Re-doped films. Temperature dependent radiative damping contributions for all Fe$_{65}$Co$_{35}$ films are given in Table 4.

Besides $\alpha_{rad}$, the $\alpha_{eddy}$ contribution should also be calculated and extracted from $\alpha_{total}$ to extract the reduced damping parameter. $\alpha_{eddy}$ can be estimated by the expression[23]

$$\alpha_{eddy} = \frac{C\gamma\mu_0^2 M_s \delta^2}{16\rho}, \qquad (6)$$

where $C$ is a parameter describing the distribution of eddy currents within the films and its value is 0.5 in our studied samples and $\rho$ is the resistivity of the films. Resistivity is measured for all films with different dopant concentrations at different temperatures. It is in the range of $8.2\times10^{-8}$ to $5.6\times10^{-8}$ $\Omega m$ for undoped, $5.7\times10^{-7}$ to $5.3\times10^{-7}$ $\Omega m$ for 3.0 at% doped, $6.9\times10^{-7}$ to $6.1\times10^{-7}$ $\Omega m$ for 6.6 at% doped and $3.9\times10^{-7}$ to $3.6\times10^{-7}$ $\Omega m$ for 12.6 at% doped films. Temperature dependent eddy current damping contributions, which are negligible, for all Fe$_{65}$Co$_{35}$ films are given in Table 5.

$\alpha_{tot}$ (filled symbols) and $\alpha_{red}$ (open symbols) versus temperature for the differently Re-doped Fe$_{65}$Co$_{35}$ films are shown in Fig. 6. Both damping parameters slowly decrease with decreasing temperature. Moreover, the damping parameter increases with increasing Re concentration; the damping parameter is 4 times as large for the 12.6 at% Re-doped sample compared to the undoped sample. Since the damping parameter depends both on disorder induced scattering and spin-orbit coupling, the observed enhancement of the damping parameter can emerge from the electronic structure of the alloy and large spin-orbit coupling of Re.

A comparison between temperature dependent experimental $\alpha_{tot}$ and $\alpha_{red}$ values and theoretically calculated intrinsic damping parameters is shown in Fig. 7 for the undoped and 12.6 at% Re-doped Fe$_{65}$Co$_{35}$ films. In agreement with the experimental results, the theoretically calculated damping parameters decrease in magnitude with decreasing temperature. It has been argued by Schoen *et al*.,[42] that the contribution to the intrinsic Gilbert damping parameter comes primarily from the strong electron-phonon coupling at high temperatures due to interband transition whereas at a low temperature, density of states at Fermi level ($n(E_F)$) and spin-orbit coupling give rise to intraband transition. In Fig. 8, we show the correspondence between the calculated damping parameter at 10 K with the density of states (spin up + spin down) at Fermi level for varying Re concentration. The increasing trend in both properties is obviously seen. The increase in DOS mainly comes from increasing DOS at Re sites in the spin-up channel. In the inset, the calculated spin-polarization as a function of Re concentration is shown. Spin polarization is defined as $\zeta = \frac{n(E_F)^\uparrow - n(E_F)^\downarrow}{n(E_F)^\uparrow + n(E_F)^\downarrow}$ where the contribution from both spin channels are seen. It is clearly observed that Re doping decreases the spin polarization.

One should note that a quantitative comparison between theory and experiment requires more rigorous theoretical considerations. The difference between experimental and theoretical results for the damping parameter may be explained by the incompleteness of the model used to calculate the Gilbert damping parameter by neglecting several complex scattering processes. Firstly, the effect of spin fluctuations was neglected, which in principle could be considered in the present methodology if the temperature dependent magnetization and hence information about the fluctuations of atomic moments were available from Monte-Carlo simulations. Other effects such as non-local damping and more sophisticated treatment of atomic displacements in terms of phonon self-energies[40] that may contribute to the relaxation of the magnetization in magnetic thin film materials have been neglected. Nevertheless, a qualitative agreement has been achieved where both experimental and theoretical results show that there is a significant increase of the damping parameter with increasing concentration of Re.

## 4. CONCLUSION

Static and dynamic magnetic properties of rhenium doped $Fe_{65}Co_{35}$ thin films have been investigated and clarified in a combined experimental and theoretical study. Results from first principles theoretical calculations show that the saturation magnetization gradually decreases with increasing Re concentration, from 2.3T for the undoped sample to 1.95T for the 10% Re-doped sample. The experimental results for the dependence of the saturation magnetization on the Re-doping are in agreement with the theoretical results, although indicating a more pronounced decrease of the saturation magnetization for the largest doping concentrations. The theoretical calculations show that the intrinsic Gilbert damping increases with increasing Re concentration; at room temperature the damping parameter is $2.8 \times 10^{-3}$, which increases to $7.3 \times 10^{-3}$ for the 10 at% Re-doped sample. Moreover, temperature dependent calculations of the Gilbert damping parameter reveal a weak decrease of the value with decreasing temperature. At a low temperature, our theoretical analysis showed the prominence of intraband contribution arising from an increase in the density of states at Fermi level. The experimental results for the damping parameter were corrected for radiative and eddy current contributions to the measured damping parameter and reveal similar trends as observed in the theoretical results; the damping parameter increases with increasing Re concentration and the damping parameter value decreases with decreasing temperature. The room temperature value for the reduced damping parameter was $2.7 \times 10^{-3}$ for the undoped sample, which increased to $9.0 \times 10^{-3}$ for the 12.6 at% Re-doped film. The possibility to enhance the damping parameter for $Fe_{65}Co_{35}$ thin films is a promising result since these materials are used in magnetic memory applications and higher data rates are achievable if the damping parameter of the material is increased.

**ACKNOWLEDGEMENT**

This work is supported by the Knut and Alice Wallenberg (KAW) Foundation, Grant No. KAW 2012.0031 and by the Marie Curie Action "Industry-Academia Partnership and Pathways" (ref. 612170, FP7-PEOPLE-2013-IAPP). The authors acknowledge financial support from Swedish Research Council (grant no. 2016-05366) and Carl Tryggers Stiftelse (grant no. CTS 12:419 and 13:413). The simulations were performed on resources provided by the Swedish National

Infrastructure (SNIC) at National Supercomputer Centre at Linköping University (NSC). M. Burak Kaynar is also acknowledged for performing resistivity measurements.

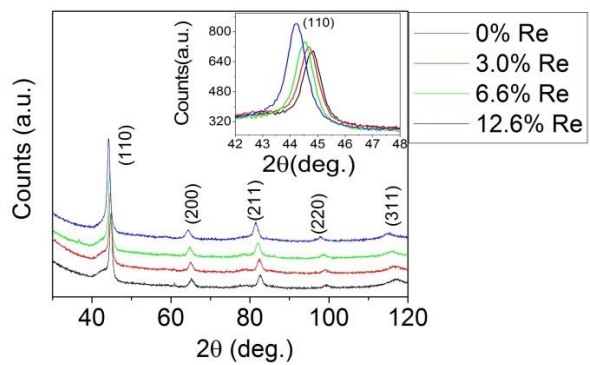 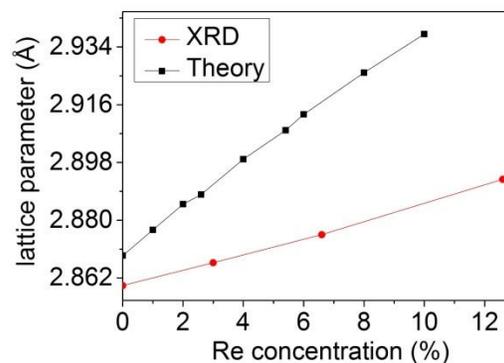

**Figure 1** (a) GIXRD plot for $Fe_{65}Co_{35}$ films with different Re concentrations. Shift of (110) peak diffraction peak with Re concentration is given as insert. (b) Lattice parameter versus Re concentration. Circles are lattice parameters extracted from XRD measurements and squares are calculated theoretical values. Lines are guide to the eye.

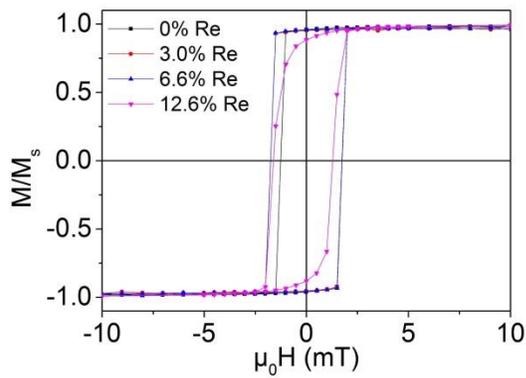 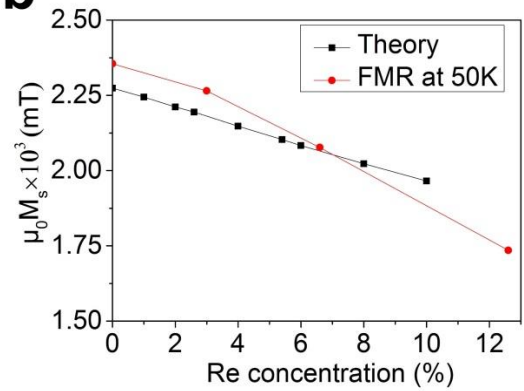

**Figure 2** (a) Normalized room temperature magnetization versus magnetic field for $Fe_{65}Co_{35}$ films with different Re concentration. (b) Low temperature saturation magnetization versus Re concentration. Circles are experimental data and squares corresponding calculated results. Experimental $\mu_0 M_s$ values were extracted from temperature dependent FMR results. Lines are guides to the eye.

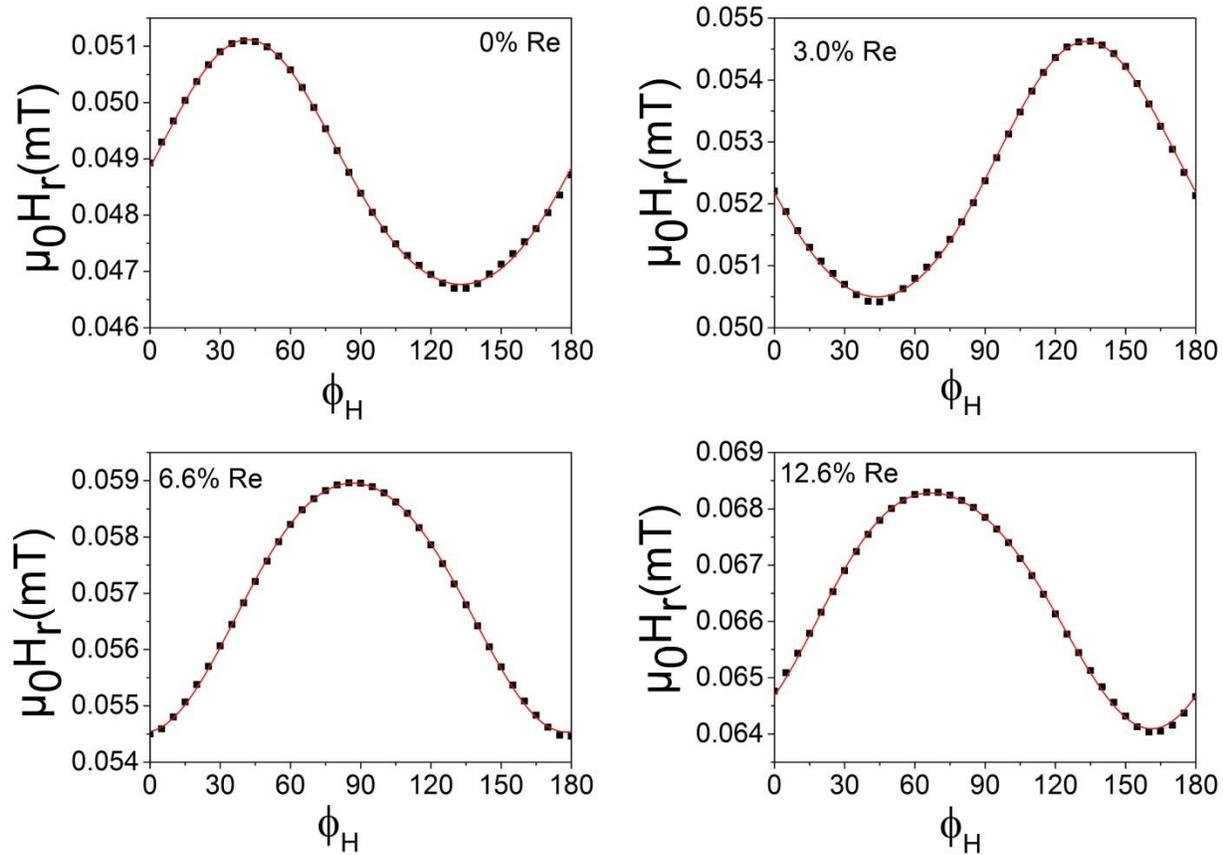

**Figure 3** $\mu_0 H_r$ versus in-plane angle of magnetic field $\phi_H$ for different dopant concentrations of Re. Black squares are experimental data and red lines are fits to Eq. (1).

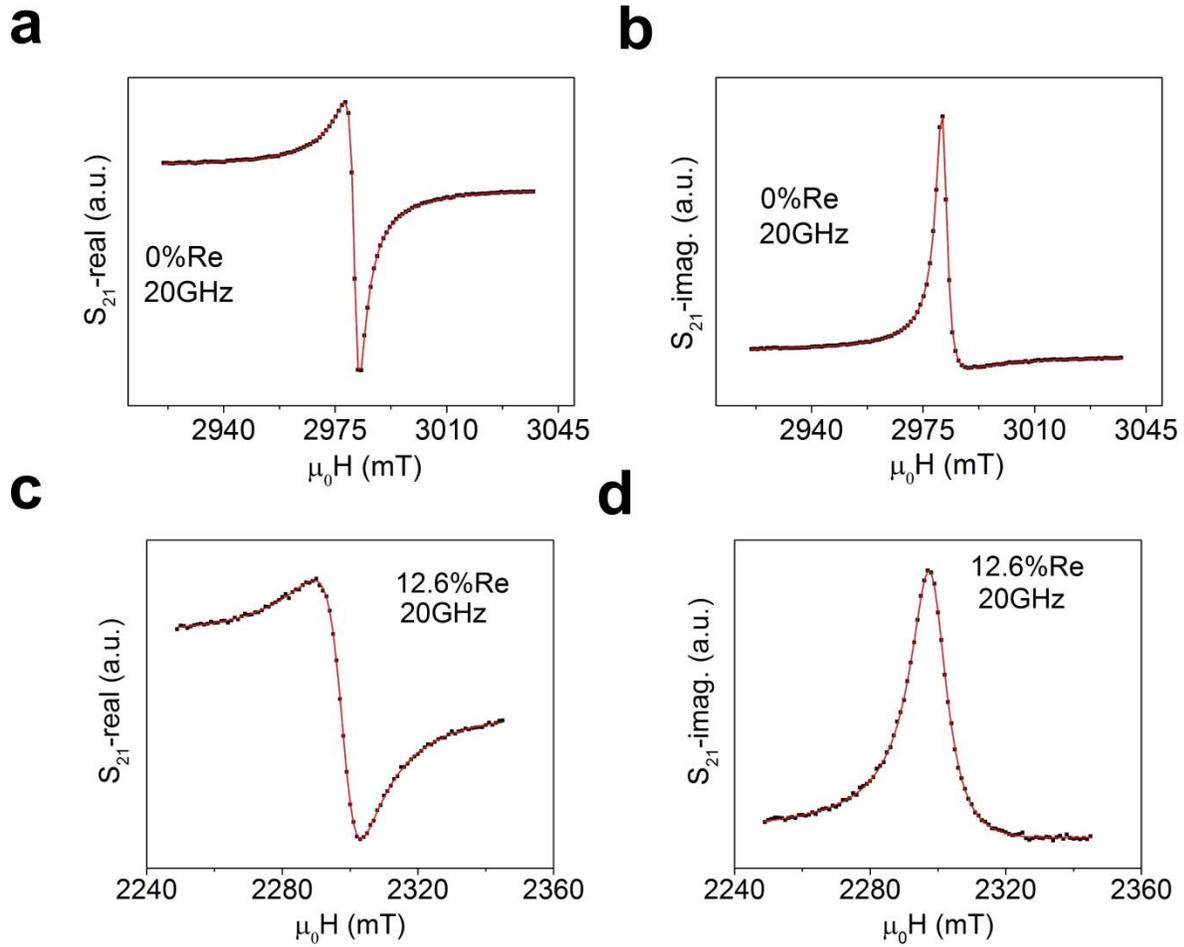

**Figure 4** Room temperature real (a and c) and imaginary (b and d) $S_{21}$ components versus out-of-plane magnetic field for $Fe_{65}Co_{35}$ thin films with 0% and 12.6 at% Re recorded at 20GHz. Black squares are data points and red lines are fits to Eq. (2).

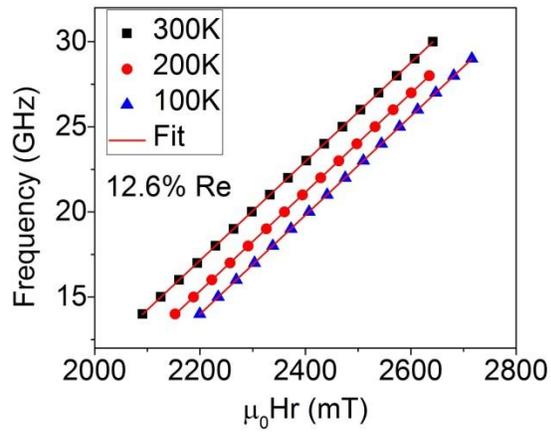 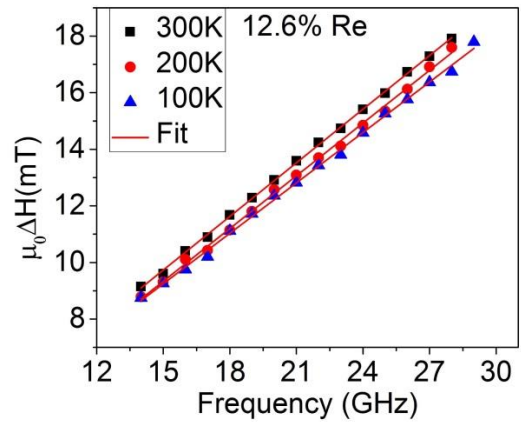

**Figure 5** (a) Frequency versus $\mu_0 H_r$ values at different temperatures for the $Fe_{65}Co_{35}$ thin film with 12.6 at% Re. Coloured lines correspond to fits to Eq. (4). (b) Linewidth $\mu_0 \Delta H$ versus frequency at different temperatures for the same Re doping concentration. Coloured lines correspond to fits to Eq. (3). Symbols represent experimental data.

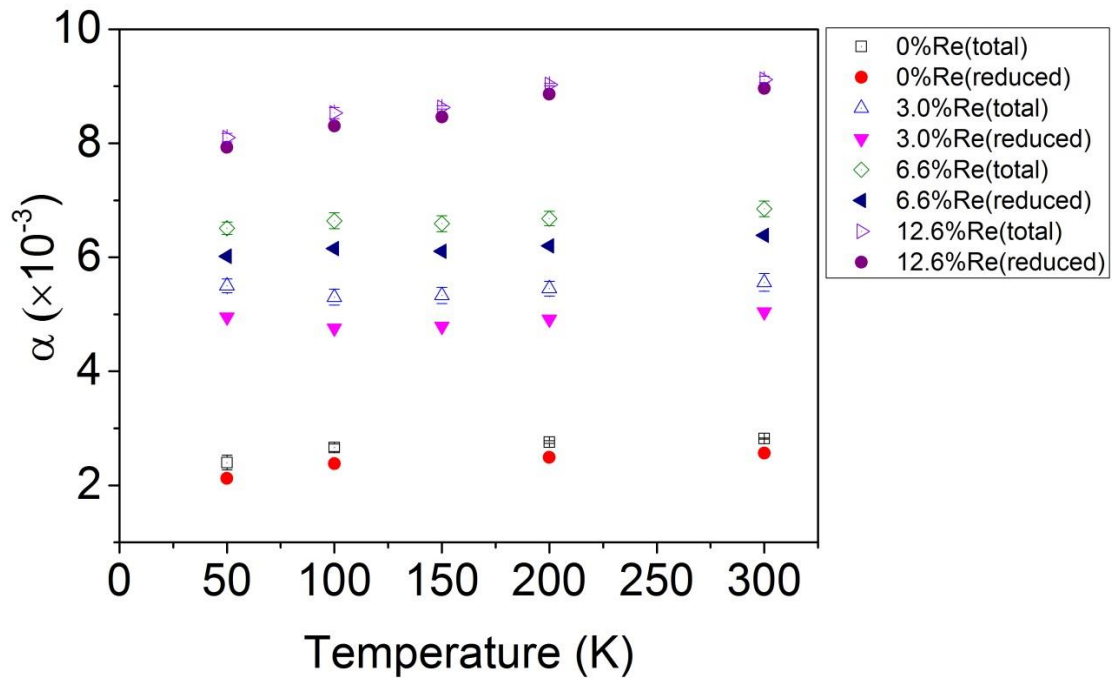

**Figure 6** $\alpha_{tot}$ versus temperature for $Fe_{65}Co_{35}$ thin films with different concentration of Re. Besides showing $\alpha_{tot}$, reduced $\alpha_{red}$ values are also plotted obtained by subtraction of radiative damping and eddy current damping contributions from $\alpha_{tot}$. Error bars are given for measured $\alpha_{tot}$ (same size as symbol size).

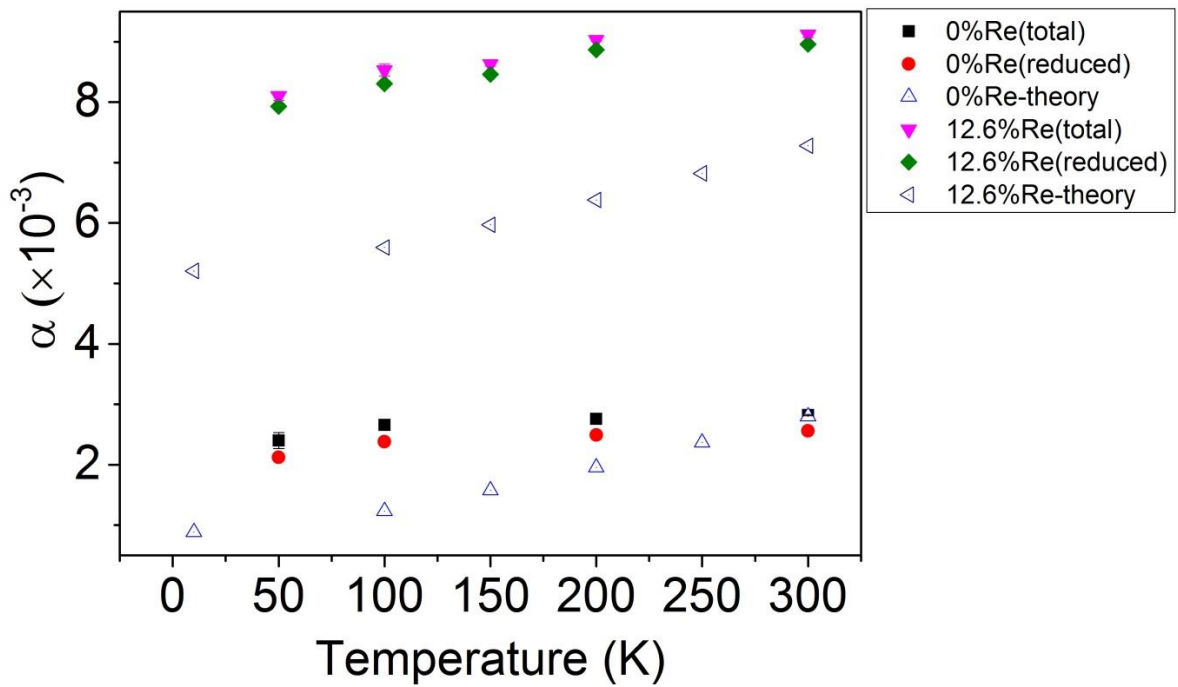

**Figure 7** $\alpha_{tot}$ versus temperature for $Fe_{65}Co_{35}$ thin films with 0 at% and 12.6 at% concentration of Re. Besides showing $\alpha_{tot}$, reduced $\alpha_{red}$ values are also plotted obtained by subtraction of radiative damping and eddy current damping contributions from $\alpha_{tot}$. In addition to experimental results theoretically calculated intrinsic damping parameters are given for the similar concentrations of Re. Error bars are given for measured $\alpha_{tot}$ (same size as symbol size).

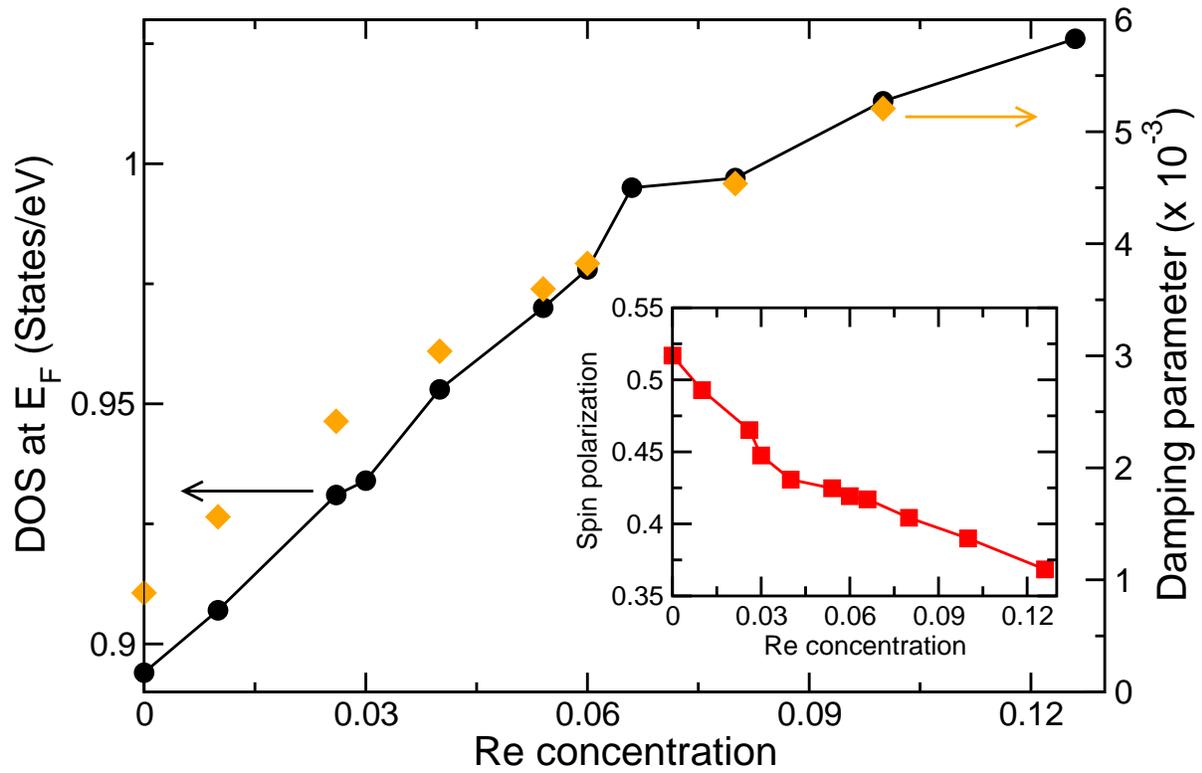

**Figure 8** Calculated density of states at Fermi level (left axis) and damping parameter (right axis) are shown as a function of Re concentration. In the inset, spin-polarization is shown as a function of Re concentration.

| Re (at%) | $t_{Ru,cap}$ (nm) | σ (nm) | $t_{FeCo}$ (nm) | σ (nm) | $t_{Ru,seed}$ (nm) | ☐ (nm) |
|---|---|---|---|---|---|---|
| 0 | 2.46 | 1.89 | 39.71 | 0.67 | 2.74 | 0.66 |
| 3.0 | 2.47 | 1.80 | 37.47 | 0.59 | 2.45 | 1.03 |
| 6.6 | 1.85 | 0.50 | 37.47 | 0.51 | 2.13 | 0.90 |
| 12.6 | 2.15 | 1.49 | 37.38 | 0.64 | 1.89 | 1.03 |

**Table 1** Thickness and roughness (σ) values for different layers in films extracted from XRR data. Error margin is ±0.02nm for all thickness and roughness values.

| Re (at%) | $\mu_0 H_u$ (mT) | $\mu_0 M_{eff}$ (T) |
|---|---|---|
| 0 | 2.20 | 2.31 |
| 3.0 | 2.10 | 2.12 |
| 6.6 | 2.30 | 1.95 |
| 12.6 | 2.20 | 1.64 |

**Table 2** Room temperature $\mu_0 M_{eff}$ and $\mu_0 H_u$ values for Fe$_{65}$Co$_{35}$ films with different concentration of Re extracted by fitting the angle dependent cavity FMR data to Eq. (1).

| Temperature (K) | 0% Re $\mu_0 M_{eff}$ (T) | 3.0 at% Re $\mu_0 M_{eff}$ (T) | 6.6 at% Re $\mu_0 M_{eff}$ (T) | 12.6 at% Re $\mu_0 M_{eff}$ (T) |
|---|---|---|---|---|
| 300 | 2.29 | 2.16 | 1.99 | 1.61 |
| 200 | 2.31 | 2.16 | 2.04 | 1.67 |
| 150 | 2.33 | 2.24 | 2.06 | 1.70 |
| 100 | 2.36 | 2.25 | 2.07 | 1.72 |
| 50 | 2.36 | 2.27 | 2.08 | 1.74 |

**Table 3** Temperature dependent $\mu_0 M_{eff}$ values for Fe$_{65}$Co$_{35}$ films with different concentration of Re extracted by fitting broadband out-of-plane FMR data to Eq. (4). Error margin is about 10 mT.

| Temperature(K) | $\alpha_{rad}$ (×10$^{-3}$) | | | |
|---|---|---|---|---|
| | 0% Re | 3.0 at% Re | 6.6 at% Re | 12.6 at% Re |
| 300 | 0.218 | 0.482 | 0.438 | 0.154 |
| 200 | 0.222 | 0.494 | 0.450 | 0.160 |
| 150 | 0.216 | 0.499 | 0.454 | 0.162 |
| 100 | 0.225 | 0.502 | 0.456 | 0.219 |
| 50 | 0.221 | 0.505 | 0.458 | 0.166 |

**Table 4** Temperature dependent radiative damping contribution to total damping parameter for Fe$_{65}$Co$_{35}$ films with different concentration of Re calculated using Eq. (5).

| Temperature(K) | $\alpha_{eddy}$ (×10⁻³) | | | |
| --- | --- | --- | --- | --- |
| | 0% Re | 3.3 at% Re | 6.6 at% Re | 12.6 at% Re |
| 300 | 0.038 | 0.077 | 0.064 | 0.006 |
| 200 | 0.047 | 0.081 | 0.067 | 0.006 |
| 150 | 0.050 | 0.084 | 0.070 | 0.006 |
| 100 | 0.055 | 0.084 | 0.073 | 0.007 |
| 50 | 0.058 | 0.086 | 0.075 | 0.007 |

**Table 5** Temperature dependent eddy current damping contribution to total damping parameter for $Fe_{65}Co_{35}$ films with different concentration of Re calculated using Eq. (6).

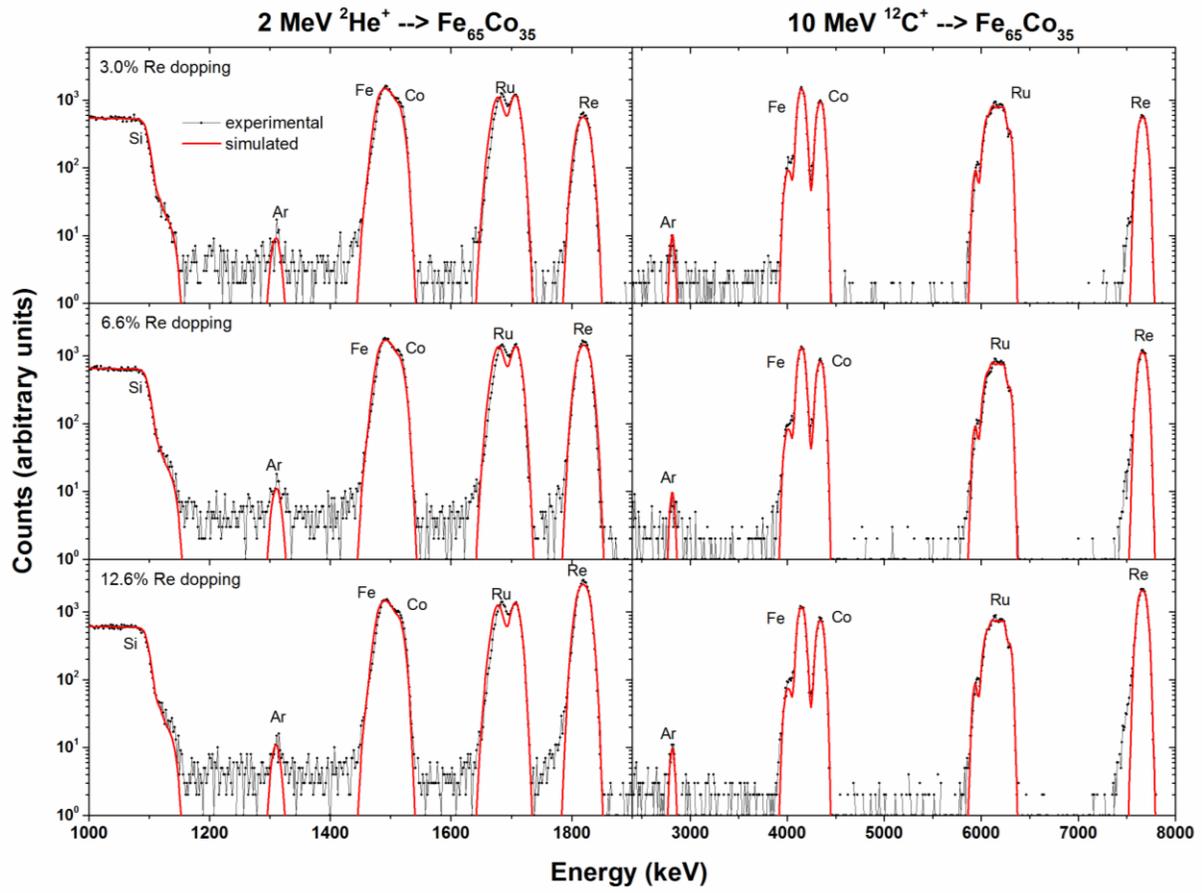

**Figure A1** RBS spectra for the Re-doped $Fe_{65}Co_{35}$ films.